\documentclass[11pt]{amsart}
\usepackage{times,amssymb,amsmath,float} 
\usepackage{euscript}
\usepackage[dvips]{epsfig}

\evensidemargin \oddsidemargin
\newtheorem{theorem}{Theorem}
\newtheorem{lemma}[theorem]{Lemma}

\newtheorem{corollary}[theorem]{Corollary}

\renewenvironment{proof}[1]{\noindent {\it Proof~:} #1}
{\ \rule{1mm}{2mm}\medskip}

\newcommand{\remove}[1]{}

\newcommand\wt{\mbox{{\rm wt}\,}}
\def\supp{\qopname\relax{no}{supp}}

\def\wt{\qopname\relax{no}{w}}

\newcommand\nc\newcommand
\nc\bfa{{\mathbf a}}\nc\bfA{{\mathbf A}}\nc\cA{{\mathcal A}}
\nc\bfb{{\mathbf b}}\nc\bfB{{\mathbf B}}\nc\cB{{\mathcal B}}
\nc\bfc{{\mathbf c}}\nc\bfC{{\mathbf C}}\nc\cC{{\mathcal C}}
\nc\bfd{{\mathbf d}}\nc\bfD{{\mathbf D}}\nc\cD{{\mathcal D}}
\nc\bfe{{\mathbf e}}\nc\bfE{{\mathbf E}}\nc\cE{{\mathcal E}}
\nc\bff{{\mathbf f}}\nc\bfF{{\mathbf F}}\nc\cF{{\mathcal F}}
\nc\bfg{{\mathbf g}}\nc\bfG{{\mathbf G}}\nc\cG{{\mathcal G}}
\nc\bfh{{\mathbf h}}\nc\bfH{{\mathbf H}}\nc\cH{{\mathcal H}}
\nc\bfi{{\mathbf i}}\nc\bfI{{\mathbf I}}\nc\cI{{\mathcal I}}
\nc\bfj{{\mathbf j}}\nc\bfJ{{\mathbf J}}\nc\cJ{{\mathcal J}}
\nc\bfk{{\mathbf k}}\nc\bfK{{\mathbf K}}\nc\cK{{\mathcal K}}
\nc\bfl{{\mathbf l}}\nc\bfL{{\mathbf L}}\nc\cL{{\mathcal L}}
\nc\bfm{{\mathbf m}}\nc\bfM{{\mathbf M}}\nc\cM{{\mathcal M}}
\nc\bfn{{\mathbf n}}\nc\bfN{{\mathbf N}}\nc\cN{{\mathcal N}}
\nc\bfo{{\mathbf o}}\nc\bfO{{\mathbf O}}\nc\cO{{\mathcal O}}
\nc\bfp{{\mathbf p}}\nc\bfP{{\mathbf P}}\nc\cP{{\mathcal P}}
\nc\bfq{{\mathbf q}}\nc\bfQ{{\mathbf Q}}\nc\cQ{{\mathcal Q}}
\nc\bfr{{\mathbf r}}\nc\bfR{{\mathbf R}}\nc\cR{{\mathcal R}}
\nc\bfs{{\mathbf s}}\nc\bfS{{\mathbf S}}\nc\cS{{\mathcal S}}
\nc\bft{{\mathbf t}}\nc\bfT{{\mathbf T}}\nc\cT{{\mathcal T}}
\nc\bfu{{\mathbf u}}\nc\bfU{{\mathbf U}}\nc\cU{{\mathcal U}}
\nc\bfv{{\mathbf v}}\nc\bfV{{\mathbf V}}\nc\cV{{\mathcal V}}
\nc\bfw{{\mathbf w}}\nc\bfW{{\mathbf W}}\nc\cW{{\mathcal W}}
\nc\bfx{{\mathbf x}}\nc\bfX{{\mathbf Z}}\nc\cX{{\mathcal X}}
\nc\bfy{{\mathbf y}}\nc\bfY{{\mathbf Y}}\nc\cY{{\mathcal Y}}
\nc\bfz{{\mathbf z}}\nc\bfZ{{\mathbf Z}}\nc\cZ{{\mathcal Z}}
\nc\od{{\bar d}}\nc\ow{{\bar w}}\nc\odelta{{\bar\delta}}
\nc\ox{{\bar x}}\nc\oy{{\bar y}}\nc\ou{{\bar u}}
\nc\oh{{\bar h}}

\nc\dgv{\delta_{\text{\rm GV}}}
\nc\rcrit{R_{\text{\rm crit}}}
\nc\Esp{E_{\text{\rm sp}}}
\renewcommand\epsilon{\varepsilon}

\newcommand{\beeq}{\begin{eqnarray*}}
\newcommand{\eneq}{\end{eqnarray*}}

\newcommand{\bigo}{\EuScript O}
\newcommand{\aux}{{\rm aux}}

\begin{document}
\title[Distance of expander codes]
{Distance properties of expander codes} 
\author[A. Barg]{Alexander Barg$^\ast$}
\thanks{$^\ast$  Supported in part by NSF grant CCR 0310961.}
\address{Dept. of ECE, University of Maryland, College Park, MD 20742}
\email{abarg@ieee.org}
\author[G. Z\'emor]{Gilles Z\'emor}
\address{\'Ecole Nationale Sup\'erieure des 
T\'el\'ecommunications, 46 rue Barrault,
75 634 Paris 13, France} \email {zemor@enst.fr}

\begin{abstract}
We study the minimum distance of codes defined on bipartite graphs.
Weight spectrum and the minimum distance of a random ensemble of
such codes are computed. It is shown that if the vertex codes have minimum
distance $\ge 3$, the overall code is asymptotically good, and sometimes
meets the Gilbert-Varshamov bound. 

Constructive families of expander codes are presented whose minimum
distance asymptotically exceeds the product bound for all code rates
between 0 and 1.
\end{abstract}
\vskip-1cm 
\maketitle

\vspace*{-1cm}

\section{Introduction}
\subsection{Context}
The general idea of constructing codes on graphs first appeared in
Tanner's classical work \cite{tan81}. One of the methods put forward in
this paper was to associate message bits
with the edges of a graph and use
a short linear code as a local
constraint on the neighboring edges of each vertex. 
M. Sipser and D. Spielman~\cite{sip96} generated
renewed
interest in this idea by tying spectral properties of the graph 
to decoding analysis of the associated code: they suggested the term
{\em expander codes} for code families whose analysis relies on graph
expansion. Further studies of 
expander codes include \cite{zem01,gur02,bar02c,bar02g,bar03a,ska03}. 

While \cite{tan81} and \cite{sip96} did not especially favor the choice of 
an underlying bipartite graph,
subsequent papers, starting with \cite{zem01}, made heavy use of this
additional feature.
In retrospect, codes on bipartite graphs can
be viewed as a natural generalization of R.~Gallager's low-density 
parity-check codes.
Another view of bipartite-graph codes involves the so-called
{\em parallel concatenation} of codes which refers to the fact that
message bits enter two or more unrelated sets of parity-check
equations that correspond to the local constraints. This view ties
bipartite-graph codes to turbo codes and related code families; 
the bipartite graph can be defined by a permutation of message
symbols which is very close to the ``interleaver'' of the turbo coding 
schemes.

A more traditional method of code concatenation, dating back to the 
classical works of P.~Elias and G.~D.~Forney, suggests to encode the message
by several codes successively, earning this class of constructions
the name {\em serial concatenation}. A well-known set of results
on constructions, parameters and decoding performance of serial
concatenations includes Forney's bound on the error exponent
attainable under a polynomial-time decoding algorithm \cite{for66},
implying in particular the existence of a constructive 
capacity-achieving code family,
and the Zyablov bound on the relative distance attainable under the
condition of polynomial-time constructibility \cite{zya71}.
Initial results of this type for expander codes \cite{sip96,zem01,bar02c} 
were substantially
weaker than both the Forney and Zyablov
bounds, but additional ideas employed both in code construction
and decoding led to establishing these results for the class of
expander codes \cite{bar03a,gur02,ska03}.
In particular, paper \cite{bar03a} focussed on similarities and 
differences between serial concatenations and bipartite-graph codes 
viewed as parallel concatenations. We refer to this paper for
a detailed introduction to properties of both code families. Paper
\cite{bar03a} also suggested a decoding algorithm 
that corrects a fraction of errors approaching half the designed
distance, i.e. half the Zyablov bound.
The error exponent of this algorithm reaches the Forney
bound for serial concatenations. The advantage of bipartite-graph
codes over the latter is that for them, the decoding complexity is an order
of magnitude lower (proportional to the block length
$N$ as opposed to $N^2$ for serial concatenations).

The main goal of \cite{bar03a} was to 
catch up with the
classical achievements of serial concatenation and show that they
can be reproduced by parallel schemes, 
with the added value of lower-complexity
decoding. One of the motivations for the
present paper is to exhibit new achievements of parallel
concatenation, unrelated to decoding, that surpass the present-day
performance of all codes constructed
in the framework of the classical serial approach.

\subsection{Bounding the minimum distance of expander codes}
The main focus of this paper are the parameters $[N,RN,\delta N]$
of bipartite-graph codes, particularly the asymptotic
behavior of the relative minimum distance $\delta$ as a
function of the rate $R$. Bipartite-graph codes, and more generally
codes defined on graphs, are famous for their low-complexity decoding
and its performance under high noise, but are generally considered to
have poorer minimum distances than their algebraic counterparts.
We strive here to reverse this trend and show that it is possible to
design codes defined on graphs with very respectable $R$ versus 
$\delta$ tradeoffs.

In the first half of this paper we study the average weight 
distribution of the random ensemble of bipartite-graph codes 
(Section \ref{sect:random}). Under the assumption that the minimum
distance of the small constituent codes
is at least~$3$, we show that the 
ensemble contains codes which are asymptotically good for all code 
rates, and for some values of the rate reach the Gilbert-Varshamov 
(GV) bound. 
This result shows 
interesting parallels with a similar theorem for serial concatenations 
in Forney's sense \cite{blo82,tho83}. It also
generalizes the result of \cite{bou99,len99} where 
bipartite-graph codes with component Hamming codes were shown to be 
asymptotically good. 

In the second part of the paper  we 
turn to constructive issues. Until now the product of the relative
distances of the constituent codes was the standard lower bound
on the relative minimum distance of expander codes,
as it is for the class of Forney's serially concatenated codes,
including product codes. Efforts have been made to surpass this
product bound, or designed distance, for short block lengths, see e.g.
\cite{tan02}, but no asymptotic improvements have been obtained for
any of these classes. In Section \ref{sect:improved}
we describe two families of 
bipartite-graph codes that asymptotically surpasses the product bound
on the minimum distance. In particular we obtain a polynomially
constructible family of binary codes that for any rate between 0 and~1
have relative distance greater than the Zyablov bound \cite{zya71}.
These constructions are based on allowing both binary and nonbinary
local codes in the expander code construction and matching the
restrictions imposed by them on the binary weight of the edges in the
graph. 
This result confirms the intuition, supported by examples of short codes
and ensemble-average results, of the product bound being
a poor estimate of the true distance of two-level code constructions
be they parallel or serial concatenations. Even though it does not
match the distance of such code families as multilevel concatenations
or serial concatenations with algebraic-geometry outer codes, this
result is still the first of its kind because all the other constructions
rely on the product bound for estimating the designed distance.
In particular, the results of Section \ref{sect:improved} improve over 
the parameters
of all previously known polynomial-time constructions of expander codes and of
concatenations of two codes not involving algebraic-geometry codes,
including the constructions of Forney \cite{for66,zya71},
Alon et al.~\cite{alo92}, Sipser and Spielman \cite{sip96}, Guruswami and 
Indyk \cite{gur02}, the authors \cite{bar02c}, and Bilu and 
Hoory \cite{bil04}. In the final Section~\ref{sect:complexity} we compare 
construction complexity with  other code families
whose parameters are comparable to those of the bipartite-graph codes
constructed in this paper.

\section{Preliminaries}\label{sect:prelim}
\subsection{Bipartite-graph codes: Basic construction}\label{sec:basic}
Let $G=(V,E)$ be a balanced, $\Delta$-regular bipartite graph with
the vertex set $V=V_0\cup V_1, |V_0|=|V_1|=n. $ 
The number of edges is $|E|=N=\Delta n$.

Let us choose an arbitrary ordering of edges of the
graph which will be fixed throughout the construction.
For a given vertex $v\in G$ this defines an ordering of edges 
$v(1),v(2),\dots,v(\Delta)$ incident to it. We denote this subset of
edges by $E(v)$. For a vertex $v$ in one part of $G$ the set of vertices 
in the other part adjacent to $v$ will be also called the 
{\em neighborhood} of the vertex $v$, denoted $\cN(v)$.

Let $A[\Delta,R_0\Delta],B[\Delta,R_1\Delta]$ be binary linear codes.
The binary bipartite-graph code $C(G;A,B)$ has parameters $[N,RN]$. We
assume that the coordinates of $C$ are in one-to-one correspondence
with the edges of $G$. Let $\bfx\in\{0,1\}^N$. By $\bfx_v$ we denote
the projection of $\bfx$ on the edges incident to $v.$
By definition, $\bfx$ is a codevector of $C$ if 
\begin{enumerate}
  \item for every $v\in V_0$, the vector $\bfx_v$ is a codeword of $A$;
  \item for every $w\in V_1$, the vector $\bfx_w$ is a codeword of $B$.
\end{enumerate}
This construction and its generalizations are primarily studied in
the asymptotic context when $n\to \infty, \Delta=\text{const}.$
Paper \cite{sip96} shows that, for a suitable choice of the code $A$, 
codes $C(G;A,A)$ are asymptotically good and
correct a fraction of errors that grows linearly with $n$ under a 
linear-time decoding algorithm. Another decoding algorithm, which 
gives a better estimate of the number of correctable errors, was suggested
in \cite{zem01}. Paper \cite{bar02c} shows that introducing two different 
codes $A$ and $B$ enables one to prove that the codes $C(G;A,B)$
attain capacity of the binary symmetric channel. 
Note that taking $A$ the parity-check code and $B$ the repetition
code we obtain Gallager's LDPC codes. For this reason codes
$C(G;A,B)$ are sometimes called generalized low-density codes 
\cite{bou99,len99}.

Before turning to parameters of the code $C$ let us recall some
properties of the graph $G$. Let $\lambda$ be the second largest eigenvalue
(of the adjacency matrix) of $G$. For a vertex $v\in V_0$ and a subset
$T\subset V_1$ let $\deg_T(v)$ be the number of edges that connect
$v$ to vertices in $T$. A key tool for the analysis of the code
$C$ is given by the following lemma. 
\begin{lemma}\cite{bar03a}\label{lemma:expanding} 
Let $S\subset V_0, T\subset V_1.$ Suppose that 
   $$
      \forall_{v\in S} \deg_T(v)\ge \alpha_0 \Delta,
  \quad \forall_{w\in T} \deg_S(w)\ge \alpha_1 \Delta,
   $$
where $\alpha_0,\alpha_1\in (0,1).$
Then
   $$
     |S|\ge \alpha_1 n \Big(1-\frac{\lambda}{\Delta\alpha_0}\Big)
                        \Big(1-\frac{\lambda}{2\Delta\alpha_1}\Big).
   $$
\end{lemma}  
From this, the relative distance of $C$ satisfies
  \begin{equation}\label{eq:distance}
    \delta\ge \delta_0\delta_1\Big(1-\frac\lambda{d_0}\Big)
                  \Big(1-\frac\lambda{2d_1}\Big),
  \end{equation}
where $d_0=\Delta\delta_0, d_1=\Delta\delta_1$ are the distances of
the codes $A$ and $B$.

The rate of the code $C$ is easily estimated to be
  \begin{equation}\label{eq:rate}
     R\ge R_0+R_1-1.
  \end{equation}

We will assume that the second eigenvalue $\lambda$
of the graph $G_1=(V_0\cup V_1,E_1)$ is small compared to its degree 
$\Delta_1$. For instance, the graph $G_1$ 
can be chosen to be Ramanujan, i.e., $\lambda\le 2\sqrt{\Delta_1-1}$. Then 
from \ref{eq:distance} we see that the code $C$ approaches the product
bound $\delta_0\delta_1$ which is a standard result for serial
concatenations.

\subsection{Multiple edges}\label{sect:mult}
In \cite{bar02c} this construction was generalized by allowing every
edge to carry $t$ bits of the codeword instead of just one bit, where
$t$ is some constant.
\remove{The following modification of the above construction was suggested in
\cite{bar02c}. Suppose that every edge in $E$ is replaced with $t$
parallel edges, where $t$ is a constant.} 
The code length then becomes $n\Delta t$. 
We again denote this quantity by $N$ because
it will always be clear from the context which of the two constructions
we consider. Let $A[t\Delta,R_0t\Delta]$ be a binary linear code
and $B[\Delta, R_1\Delta]$ be a $q$-ary additive code, $q=2^t$.
To define the code $C(G;A,B)$ we keep Condition 1. above and replace
condition 2 with
\begin{quote}
 $2'.$ for every $w\in V_1$ the vector $\bfx_w$, {\em viewed as a $q$-ary
vector\/}, is a codeword in $B$.
\end{quote}
An alternative view of this construction is allowing $t$ parallel edges
to replace each edge in the original graph $G$. Then every edge again
corresponds to one bit of the codeword. 
An advantage of the view offered above is that it allows a direct
application of Lemma \ref{lemma:expanding}.

In \cite{bar02c} it was shown that this improves the parameters and 
performance estimates of the code $C$. For instance, there exists
an easily constructible code family $C(G;A,A)$ of rate $R$ with relative 
distance given by 
  \begin{equation}\label{eq:mult}
    \delta\ge \frac12(1-R) h^{-1}\Big(\frac{1-R}2\Big)
             -\epsilon \qquad(\epsilon>0)
  \end{equation}
where $h(\cdot)$ is the binary entropy function. Note that the distance
estimate is immediate from Lemma \ref{lemma:expanding}. 

The generalized codes of this section together with some other 
modifications of the original construction will be used in 
Sect.~\ref{sect:improved} below.

\begin{figure}[tH]
\epsfysize=12cm
\setlength{\unitlength}{1cm}
\begin{center}
\begin{picture}(5,13)
\put(-3,1){\epsffile{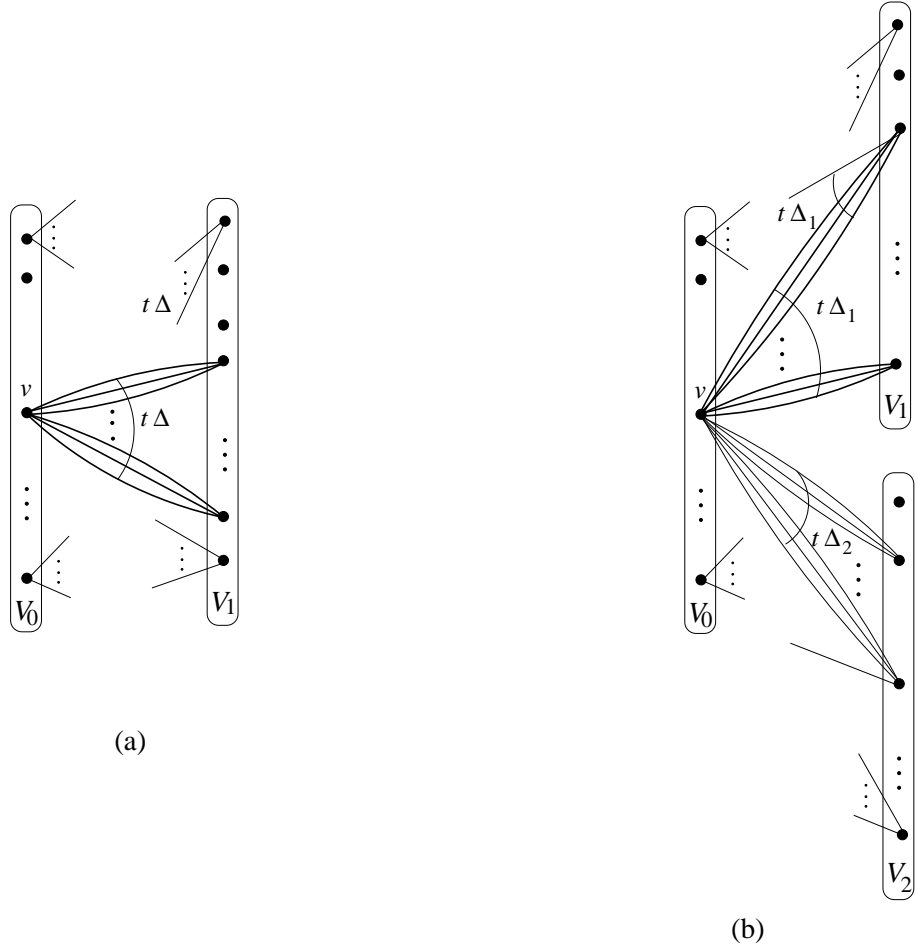}}
\put(6.5,0.5){{ (b)}}
\put(-1.7,3){{ (a)}}
\end{picture}
\vskip-5mm
\caption{Constructions of bipartite-graph codes with
multiple edges: (a) Basic construction, (b) Modified construction.}
\label{fig:code-constructions}
\end{center}
\end{figure}

\subsection{Modified code construction}\label{sect:mod}
Let $G=(V,E)$ be a bipartite 
graph whose parts are $V_0$ (the left vertices) and $V_1\cup V_2$ 
(the right vertices), where $|V_i|=n$ for $i=0,1,2.$ 
The degree of the left vertices is $\Delta$,
the degree of the vertices in $V_1$ is $\Delta_1$ and the degree
of vertices in $V_2$ is $\Delta_2=\Delta-\Delta_1.$ 
For a given vertex $v \in V_0$ 
we denote by $E(v)$ the set of all
edges incident to it and by $E_i(v)\subset E(v), i=1,2$ the subset of 
edges  of the form $(v,w)$, where $w\in V_i$. 
The ordering of the edges on
$v$ defines an ordering on $E_i(v).$ Note that both subgraphs 
$G_i=(V_0\cup V_i, E_i), i=1,2$ can be chosen to be
 regular, of degrees $\Delta_1$
and $\Delta_2$ respectively. 

Let $A$ be a $[t\Delta, R_0t\Delta,d_0=t\Delta \delta_0]$ linear binary code 
of rate $R_0=\Delta_1/\Delta$. The code $A$ can also be seen as a $q$-ary
additive $[\Delta, R_0\Delta]$ code, $q=2^t$. Let $B$ be a $q$-ary 
$[\Delta_1, R_1\Delta_1,d_1=\Delta_1\delta_1]$ additive code. 
We will also need an auxiliary $q$-ary code $A_{\text{aux}}$
of length $\Delta_1$. Every edge of the graph will be associated with
$t$ bits of the codeword of the code $C$ of length $N=nt\Delta.$
The code $C$ is defined as the set of vectors
$\bfx=\{x_1,\dots, x_N\}$ such that

\begin{enumerate}
\item For every vertex $v\in V_0$ the subvector $(x_j)_{j\in E(v)}$ is
a ($q$-ary) codeword of $A$ and the set of coordinates $E_1(v)$ is an
information set for the code $A$.
\item For every vertex $v\in V_1$ the subvector $(x_j)_{j\in E(v)}$
is a codeword of $B$;
\item For every vertex $v\in V_0$ the subvector $(x_j)_{j\in E_1(v)}$
is a codeword of $A_{\text{aux}}.$
\end{enumerate}

Both this construction and the construction from the previous subsection
are illustrated in Fig.~\ref{fig:code-constructions}.

We will choose
the minimum distance $d_{\text{aux}}=\delta_{\text{aux}}\Delta_1$ of the code 
$A_{\text{aux}}$ so as to make the quantity $\lambda/d_{\text{aux}}$
arbitrarily small, where $\lambda$ is the second eigenvalue of $G_1$.
By choosing $\Delta_1$ large enough, the rate $R_{\text{aux}}$ of
$A_{\text{aux}}$ can be thought of as a quantity such that
$1-R_{\text{aux}}$ is almost $\bigo(1/\sqrt\Delta).$

This construction was introduced and studied in \cite{bar02g,bar03a}.
The code $C$ has the parameters $[N=nt\Delta,RN,D].$
The rate $R$ is estimated easily from the construction:
  \begin{equation}\label{eq:Raux}
    R\ge R_0R_1-R_0(1-R_\aux),
  \end{equation}
which can be made arbitrarily close to $R_0R_1$ by choosing $\Delta$
large enough but finite.

The distance $D$ of the code $C$ can be again estimated from Lemma 
\ref{lemma:expanding} applied to the subgraph $G_1.$ Then we have
$\alpha_0=\delta_\aux, \alpha_1=\delta_1,$ and  
  \begin{equation}\label{eq:product}
        D \geq \delta_0\delta_1
   \left(1-\frac{\lambda}{d_\aux}\right)
             \left(1-\frac{\lambda}{2d_1}\right)N.
  \end{equation}
This means in particular that the relative minimum distance $D/N$
is bigger than a quantity that can be made arbitrarily close to
the product $\delta_0\delta_1$. Together with (\ref{eq:Raux})
this means that the distance of the code $C$ for $n\to \infty$
can be made arbitrarily close to the product, or Zyablov bound \cite{zya71}
\begin{equation}\label{eq:zyab}
    \delta_{\text Z}(R)=\max_{R\le x\le 1} \dgv(x)(1-R/x).
\end{equation}
This result was proved in \cite{bar03a}.

{\bf Alternative description} of the modified construction. The above code
can be thought of as a serially concatenated code with $A$
as inner binary code and a $Q$-ary outer code with $Q=2^{t\Delta_1}.$
The outer code is formed by viewing the binary $t\Delta_1$-tuple
indexed by the edges of $G_1$ incident to a vertex of $V_0$ as an element
of the $Q$-ary alphabet. The $Q$-ary cod $B'$ is defined by conditions
2 and 3 above, and $C$ is obtained by concatenating $B'$ with $A$. 
This description of the modified construction is used 
in \cite{ska04} to show the existence of linear-time decodable codes
that meet the Zyablov bound and attain the Forney error exponent under 
linear-time decoding on the binary symmetric channel as well as the Gaussian 
and many other communication channels.
Another closely related work is the paper \cite{gur02} 
where a similar description was used to prove that there exist 
bipartite-graph codes that meet the bound (\ref{eq:zyab}) and correct
a $\delta_{\text Z}/2$ proportion of errors under a linear-time decoding
procedure.

\section{Random ensemble of bipartite-graph codes}\label{sect:random}
Let us discuss average asymptotic properties of the ensemble
bipartite-graph codes. It has been known since Gallager's 1963 book
\cite{gal63} that
the ensemble of random low-density codes (i.e., bipartite graph codes
with a repetition code on the left and a single parity-check code on 
the right) contains asymptotically good codes whose relative distance 
is bounded away from zero for any code rate $R\in(0,1).$ 
Papers \cite{bou99} and \cite{len99} independently proved
that the ensemble of random 
bipartite-graph codes with Hamming component codes on both sides
contains asymptotically good codes.
Here we replace Hamming codes with arbitrary binary linear codes and
show that the corresponding ensemble 
contains codes that meet the GV bound.
\begin{theorem} \label{prop:R}
Let $G=(V_0\cup V_1,E)$ be a random $\Delta$-regular 
bipartite graph, $V_0=V_1=n$ and let $A[\Delta,R_0\Delta]$ be a random
linear code. 
For $n\to \infty$ the average weight distribution over the ensemble
of linear codes $C(G;A,A)$ of length $N=n\Delta$ and rate $R$ 
is bounded above as $A_{\omega N}\le 2^{NF+o(N)},$ where
    \begin{alignat}{2}
    F&= \omega[R-1- 2\log(1-2^{R_0-1})]-h(\omega) & &\quad
          \text{ if } 0<\omega\le 1-2^{R_0-1}\label{eq:sp1}\\
    F&= h(\omega)+R-1 & &\text{ if }\omega\ge 1-2^{R_0-1}\label{eq:sp2}
  \end{alignat} 
\end{theorem}
\begin{proof}
Let $H$ be a $\Delta(1-R_0)\times\Delta$ parity-check matrix of the code $A$.
The parity-check matrix of the code $C$ can be written as follows:
$\cH=[\cH_1,\cH_2]^t,$ where
\[
   \cH_1=\left[\begin{array}{cccc}
             H&&&\\
             &H&&\\
             &&{\dots}&\\
             &&&H
             \end{array}\right]
\]
(a band matrix with $H$ repeated $n$ times) and $\cH_2=\pi(\cH_1)$ is
a permutation of the columns of $\cH_1$ defined by the edges of the graph
$G$. 

To form an ensemble of random bipartite-graph codes assume that $H$ is 
a random binary matrix with uniform distribution and that the permutation $\pi$
is chosen with uniform distribution from the set of all permutation on
$N=n\Delta$ elements.
Choose the uniform probability on $Z_N=\{0,1\}^N$ and endow the product space
of couples $(\cH ,\bfx)$ with the product probability. 

The average number of codewords of weight $w$ is
\begin{equation}\label{eq:average1}
 A_w = \binom Nw \Pr[\cH \bfx^t=0\; | \; \wt(\bfx)=w].
\end{equation}
Let us compute the probability $\Pr[\cH \bfx^t=0\; | \; \wt(\bfx)=w]$.

Observe that 
\begin{eqnarray}
   \Pr[\cH \bfx^t=0\; | \; \wt(\bfx)=w] & = &
   \Pr[\cH_1\bfx^t=0\; | \; \wt(\bfx)=w]\Pr[\cH_2\bfx^t=0\; | \; \wt(\bfx)=w]
   \nonumber\\
        & = & ( \Pr[\cH_1\bfx^t=0\; | \; \wt(\bfx)=w])^2. \label{eq:square}
\end{eqnarray}

Let $w=\omega n\Delta$. 
Let $\cX_{m,w}\subset \{0,1\}^N$ be the event where $\bfx$ is of 
weight $w$ and contains nonzero
entries in exactly $m$ groups of coordinates of the form
$(x_{i\Delta+j}, j=1,\dots,\Delta; i=0,\dots,n-1)$.
Let $w_i=\omega_i\Delta$ be the number of
ones in the $i$th group. We have
  \[
  \Pr_{Z_N}[\cX_{m,w}] = 2^{-N}\binom nm\sum_{\sum w_i =w}
               \prod_{i=1}^m\binom {\Delta}{w_i}
  \cong 2^{-N}\binom nm\sum_{\sum w_i =w} 2^{\Delta \sum_ih(\omega_i)}
  \]
By convexity of the entropy function (or by using Lagrange multipliers),
the maximum of the last expression on $\omega_1,\dots,\omega_m$ under
the restriction $\sum_i\omega_i=\omega n$ is attained when 
$\omega_i=\omega n/m, i=1,\dots, m.$ 
For large $\Delta$ we therefore have
  $$ \Pr_{Z_N}[\cX_{m,w}] \cong 2^{-N +m\Delta h(\omega n/m)}$$

Now we have
  \beeq
   \Pr[\cH_1\bfx^t=0\; | \; \wt(\bfx)=w] & = & 
   \frac{\Pr[\cH_1\bfx^t=0\; ,\;\wt(\bfx)=w]}{\Pr[\wt(\bfx)=w]}\hspace{5mm}
   \text{and}\\
   \Pr[\cH_1\bfx^t=0\; ,\;\wt(\bfx)=w] & = & 
   \sum_m\Pr[\cH_1\bfx^t=0\; ,\cX_{m,w}]
  \eneq
and clearly
  $$\Pr[\cH_1\bfx^t=0\; ,\;\cX_{m,w}] = 2^{m\Delta(R_0-1)}\Pr[\cX_{m,w}]$$
so that
  $$\Pr[\cH_1\bfx^t=0\; , \; \wt(\bfx)=w] \cong 2^{-N+\max_m(m\Delta(R_0-1) +
    m\Delta h(\omega n/m))}$$
and
  $$\Pr[\cH_1\bfx^t=0\; | \; \wt(\bfx)=w] \cong 
    2^{-h(\omega)N+\max_m(m\Delta(R_0-1) + m\Delta h(\omega n/m))}.$$
Given (\ref{eq:average1}) and (\ref{eq:square}),
and setting $x = m/n$ we obtain therefore
$A_w=2^{N F(R_0,x)},$ where
  \begin{align}
   F(R_0,x)&=-h(\omega)+2\max_{\omega\le x\le 1}
            (x(R_0-1+h(\omega/x)))+o(1) \nonumber\\
       &\le -h(\omega)+\max_{\omega\le x\le 1}
            (x(R-1+2h(\omega/x)))+o(1)\label{eq:eexp}
  \end{align}
The unconstrained maximum on $x$ in the last expression is attained for 
$x=x_0=\omega/(1-z),$ where $2\log z=R-1$.
Thus, the optimizing value of $x$ equals
$x_0$ if this quantity is less that 1 and 1 otherwise.
Substituting $x=x_0$ into (\ref{eq:eexp}) and taking into account the
equality $R-1+2h(z)=2(1-z)\log(z/(1-z)),$
we obtain
  \[
    F(R_0,x)\le -h(\omega)+\frac\omega{1-z}(R-1+2h(z))
  \]
which is exactly (\ref{eq:sp1}).
Substituting $x=1$ we obtain the second part of the claim.
\end{proof}

\bigskip
The result of this theorem 
enables us to draw conclusions about the average minimum
distance of codes in the ensemble. 
From (\ref{eq:sp1})-(\ref{eq:sp2}) and the proof
it is clear that the relation between these expressions is
  \[
   \omega[R-1- 2\log(1-2^{R_0-1})]-h(\omega)\ge h(\omega)+R-1
  \]
(since $x_0$ in the previous proof is the only maximum point),
and that this inequality is strict for $\omega< 1-2^{R_0-1}.$ 
Thus if
$1-2^{R_0-1}<\dgv(R),$ the first time the exponent of the 
ensemble average weight spectrum becomes positive is $\omega=\dgv(R).$
This would mean that for large $n$ there exist codes in the bipartite-graph
ensemble that approach the GV bound; however, there is one obstacle for
this conclusion: since the exponent approaches $0$ for $\omega\to 0$
the codes in principle can contain very small nonzero weights
(such as $w$ constant or growing slower than $n$). This issue 
is addressed in the next theorem, where a slightly stronger fact is proved, 
namely that there exists a constant $\epsilon>0$ such that on average, 
the distance of the codes $C(G;A,A)$ is at least $\epsilon n.$ 
\begin{theorem}\label{th:GVbound}
Consider the ensemble of bipartite-graph
codes defined in Theorem \ref{prop:R}. Let $\omega^\ast$ be the only
nonzero root of the equation
  \[
    \omega\left(R-1-2\log(1-2^{(1/2)(R-1)})\right)=h(\omega).
  \]
The ensemble average relative distance behaves as
  \begin{alignat}{2}
    \delta(R)&= \omega^\ast & &\quad
          \text{ if } R_0\le \log(2(1-\dgv(R)))
                 \label{eq:d1}\\
    \delta(R)&= \dgv(R) & &\quad \text{ if } R_0> \log(2(1-\dgv(R)))
                 \label{eq:d2}
  \end{alignat}
In particular, for $R\le 0.202$ the ensemble contains codes that meet
the GV bound.
\end{theorem}
\begin{proof}
As argued in the discussion preceding the statement of Theorem 
\ref{th:GVbound}, we only need to check that for sufficiently small $w$, the
expected number of codewords of weight $w$ in the ensemble is a
vanishing quantity.

Suppose the local code $A$ has minimum distance $d$. Let
$w<cn$, $c$ a constant to be determined later, and set $m=w/d$.
Let $U(w,d)\subset \{0,1\}^N$ be the set of vectors with the property
that if for some $i=0,\dots,n-1$ the subvector
$(x_{i\Delta+j}, j=1,\dots,\Delta)$ is nonzero, it is of weight at least $d$.
Let $\cH_1$ be as in the proof of Theorem \ref{prop:R}.
Then
   $$
     \Pr[\cH_1 \bfx=0\mid \wt(\bfx)=w]\le \Pr[\bfx\in U(w,d) \mid \wt(\bfx)=w].
   $$
Next
  $$
    |U(w,d)|=\sum_{i=w/\Delta}^m\binom ni \binom\Delta d^i
      \binom{i\Delta}{(m-i)d}\le \binom nm\binom\Delta d^m
     \sum_{i=w/\Delta}^m \binom   {m\Delta}{(m-i)d}
  $$
  $$
       \le\binom nm\binom\Delta d^m 2^{m\Delta}.
  $$
Then
  $$
         \Pr[\cH_1 \bfx=0\mid \wt(\bfx)=w]\le
         \binom nm\binom\Delta d^m 2^{m\Delta}\binom Nw^{-1} .
  $$
Recall that $\cH_2$ is obtained by randomly permuting the columns of 
$\cH_1$, so the expected number of codewords of weight $w$ in the code
is 
  \begin{align*}
    A_w&\le \binom nm^2\binom\Delta d^{2m} 2^{2m\Delta}\binom Nw^{-1}
  \end{align*}
Remember that $\Delta$ is fixed.
Since $\binom Nw\ge N^w/w^w$, $N=\Delta n$, and $\binom nm \leq
e^m n^m/m^m$, we obtain
  \beeq
   A_w & \le & e^{2m}\frac{n^{2m}}{m^{2m}}\Delta^{2dm} 2^{2m\Delta} (w/N)^w
         \le  e^{2m}n^{m(2-d)}\frac{(\Delta w)^{dm}}{(w/d)^{2m}}2^{2m\Delta}\\
       & \le &  (s^{\frac1{2-d}} n/w)^{w(2-d)/d}
  \eneq
where $s=(ed)^2 \Delta^d 2^{2\Delta}$ is a constant 
independent of $n$. For any $w<s^{\frac1{d-2}}  n$, we get that
the right-hand side of the last inequality
tends to $0$ as $n\to\infty$ whenever $d\geq 3.$ 
\end{proof}

\begin{corollary}\label{cor:3} Consider the ensemble $\cC$ of bipartite-graph
codes defined in Theorem \ref{prop:R}. If the distance of the code
$A$ is at least $3$ then the ensemble-average relative distance is
bounded away from zero $($i.e., the ensemble $\cC$ contains asymptotically
good codes\/$)$.
\end{corollary}

\begin{figure}[tH]
\epsfysize=4.5cm
\setlength{\unitlength}{1cm}
\begin{center}
\begin{picture}(5,9)
\put(-5.5,2.5){\epsffile{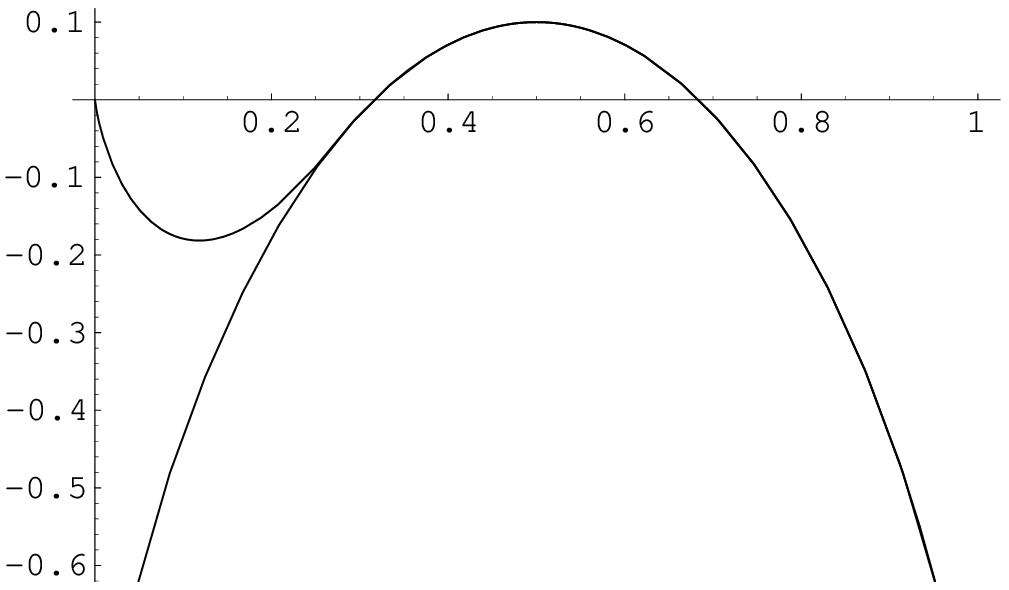}}
\put(3.5,2.5){\epsffile{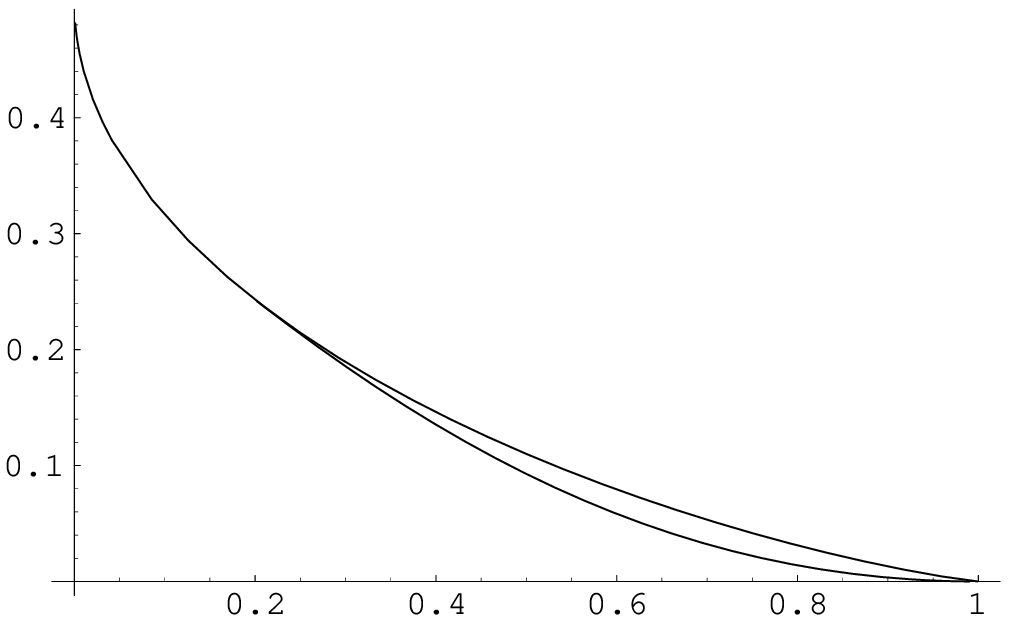}}
\put(-4.5,7.2){{\mbox{\small$N^{-1}\log A_{\omega N}$}}}
\put(2.2,6.6){{\mbox{\small$\omega$}}}
\put(4.2,6.8){{\mbox{\small$\delta$}}}
\put(10.5,2.2){{\mbox{\small$R$}}}
\put(-4.1,5.4){{\small(\ref{eq:sp1})}}
\put(-3.8,3.7){{\small(\ref{eq:sp2})}}
\put(5.1,5.4){{\small(\ref{eq:d2})}}
\put(7.1,3.1){{\small(\ref{eq:d1})}}
\put(-2.7,1.7){{ (a),} $R=0.1$}
\put(6.5,1.7){{ (b)}}
\end{picture}
\vskip-1.5cm
\caption{Average weight spectrum (a) and distance (b) of the 
ensemble of bipartite-graph codes)}\label{fig:weight-distance}
\end{center}
\end{figure}

The results of the last two theorems (ensemble-average weight spectrum and
relative distance) are shown in Fig.~\ref{fig:weight-distance}. Note that 
random bipartite-graph codes are asymptotically good for all code rates.
We also observe that the behavior of the function $\log A_{\omega N}$ is 
similar to that of the logarithm of the ensemble-average weight spectrum for
Gallager's codes (see \cite{gal63}, particularly, p. 16) and of other
LDPC code ensembles. It is interesting to note that Gallager's codes
in \cite{gal63} become asymptotically good on the average once
the number $j$ of ones in the column of the parity check matrix is 
at least 3. Similarly, we need distance-3 local codes to guarantee
relative distance bounded away from zero in the ensemble of bipartite-graph
codes.

\bigskip We conclude this section by mentioning two groups of results
related to the above theorems.

1. A different analysis of the weight spectrum of codes on graphs
with a fixed local code $A$ was performed in \cite{gal63,bou99,len99}.
Let $A$ be an $[\Delta,R_0\Delta]$ linear binary code with weight 
enumerator $a(y).$
Let  $\lambda=\lambda(\omega)$ be the root of $(\ln a(e^s))'_s=\Delta\omega$
with respect to $s$. Let $A_{\omega N}$ be the component of the 
ensemble-average weight spectrum of the code $C$. As $n\to \infty,$
we have \cite{bou99,len99}
  \begin{equation}\label{eq:hamming}
    N^{-1} \log_2 A_{\omega N}\le h(\omega) -
          \frac2{\ln 2} 
          \Big(\frac{\ln a(e^\lambda)}{\Delta} 
              - \lambda\omega\Big) +o(N).
  \end{equation}
Note that this is a Chernov-bound calculation since $a(e^s)$ is 
(proportional to) the moment generating function of the code $A$.
Variation of this method can be also used to obtain Theorem 
\ref{prop:R}, although the argument is not simpler than the direct
proof presented above. On the other hand, the proof method of 
Theorem \ref{prop:R} does not seem to lead to a closed-form expression
for the ensemble-average weight spectrum for a particular code $A$
(cf. \cite{bar01e}).

\begin{figure}[tH]
\epsfysize=4.5cm
\setlength{\unitlength}{1cm}
\begin{center}
\begin{picture}(5,9)
\put(-5.5,2.5){\epsffile{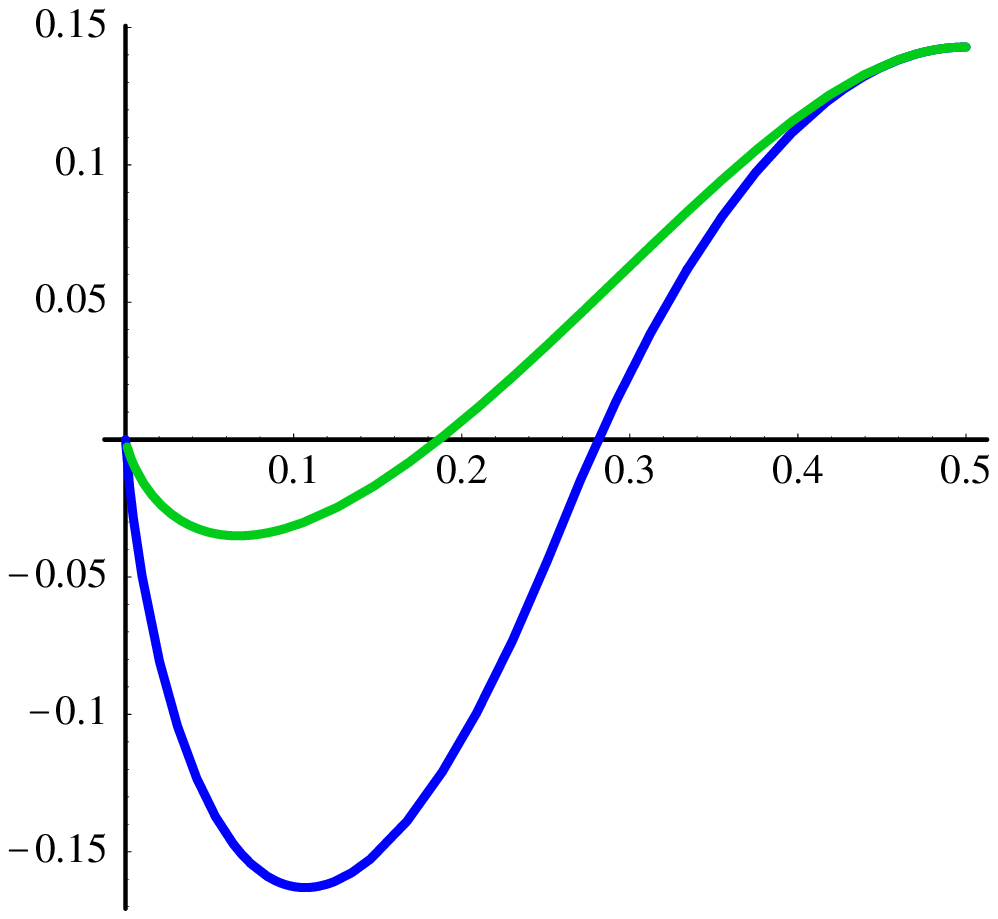}}
\put(3.5,2.5){\epsffile{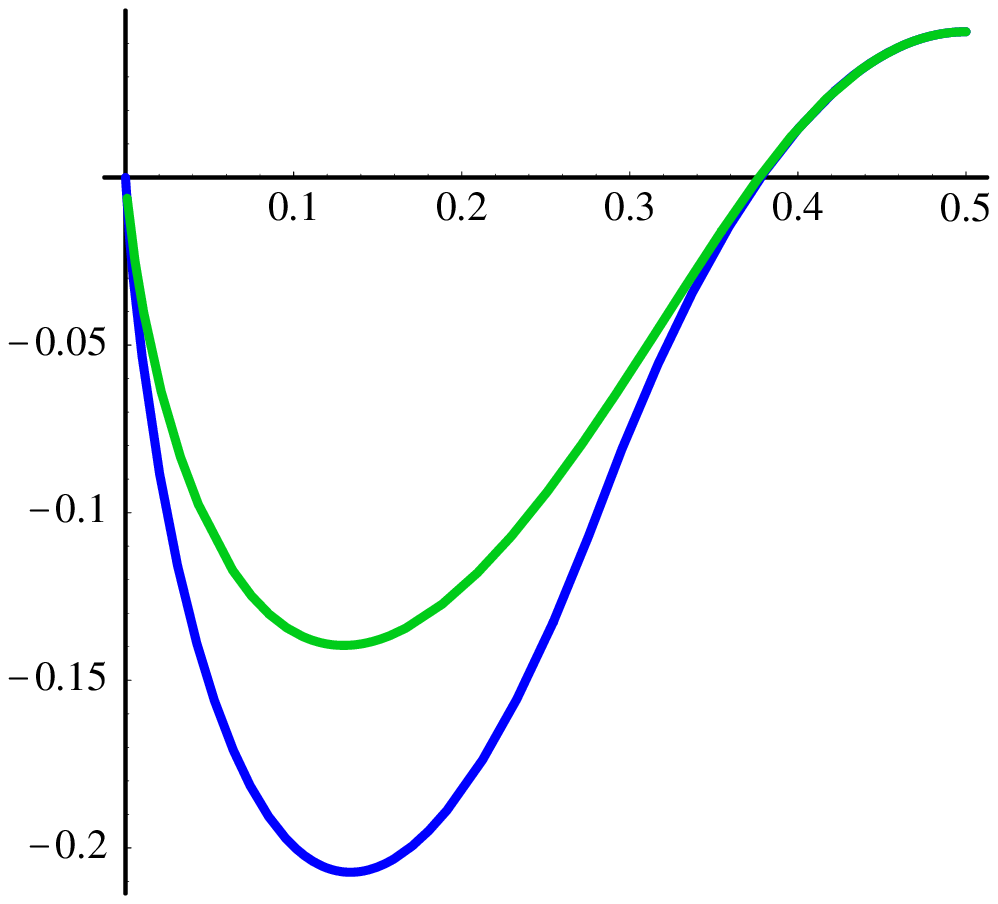}}
\put(-4.7,7.2){{\mbox{\small$N^{-1}\log A_{\omega N}$}}}
\put(4.2,7.2){{\mbox{\small$N^{-1}\log A_{\omega N}$}}}
\put(-0.8,5){{\mbox{\small$\omega$}}}
\put(8.5,6.3){{\mbox{\small$\omega$}}}
\put(-2.7,1.7){{ (a)}}
\put(6.5,1.7){{ (b)}}
\end{picture}
\vskip-1.5cm
\caption{Average weight spectrum of the code $C$\/:
(a) the local code $A$ is the Hamming code (upper curve) and a
random code (lower curve);
(b) the local code $A$ is the Golay $[23,12,7]$ code
(upper curve) and random code (lower curve).
}\label{fig:hamming}
\end{center}
\end{figure}

It is interesting to compare the weight spectrum
(\ref{eq:sp1})-(\ref{eq:sp2}) to the spectrum (\ref{eq:hamming}).
For instance let $A$ be the $[7,4,3]$ Hamming code
with $a(y)=1+7y^3+7y^4+y^7$. We plot the spectrum of the code $C$ in 
Fig.\ref{fig:hamming}(a) together with the weight spectrum 
(\ref{eq:sp1})-(\ref{eq:sp2})
and do the same for $A$ the $[23,12,7]$
Golay code in Fig.\ref{fig:hamming}(b).
For the code $C$ with local Hamming codes the parameters are: $R\ge 1/7,
\delta\ge 0.186.$ The GV distance $\dgv(1/7)\approx 0.281.$
For the case of the Golay code we have $R\ge 1/23,\delta\ge 0.3768.$
The GV distance in this case is $\dgv(1/23)\approx0.3788.$

The main result of \cite{bou99,len99} is that bipartite-graph 
codes with Hamming local codes
are asymptotically good. We remark that this also follows
as a particular case of Corollary \ref{cor:3} above.

\bigskip
2. Recall the asymptotic
behavior of other versions of concatenated codes, in particular
serial concatenations. Consider the ensemble of concatenated codes 
 with random $[\Delta,R_0\Delta]$
inner codes $A$ and MDS outer codes $B$. The following results are due
to E.~L.~Blokh and V.~V.~Zyablov \cite{blo82} and C.~Thommesen \cite{tho83}.
The average weight spectrum
is given by $A(N\omega)=2^{N(F+o(1))},$ where
  \begin{alignat*}{2}
   F &=R-R_0-\omega\log(2^{1-R_0}-1) &&\qquad0<\omega\le 1-2^{R_0-1}\\
   F &= h(\omega)+R-1 &&\qquad\omega\ge 1-2^{R_0-1}.
  \end{alignat*}
The ensemble average relative
distance is given by 
 \begin{equation*}
   \delta(R)=\begin{cases}
          \dgv(R) &R_0\ge \log(2(1-\dgv(R)))\\[2mm]
          \frac{R-R_0}{\log(2^{1-R_0}-1)} &0\le R_0< \log(2(1-\dgv(R)))
        \end{cases}
 \end{equation*}
where $R$ is the code rate. 
With all the similarity of these results to those proved in this
section there is one substantial difference: with serial concatenations
there is full freedom in choosing the rate $R_0$ of the inner codes
while with parallel codes once the overall rate is fixed the
rate $R_0$ of the code $A$ is also fully determined. This explains
the fact that serially concatenated codes meet the GV bound for
all rates $R$ while bipartite-graph codes do so only for relatively
low code rates.

Another result worth mentioning in this context \cite{bar01e} 
concerns behavior of serially concatenated codes with a {\em fixed} 
inner code $A$ and random outer $q$-ary code $B$.
This ensemble can be viewed as a serial version of the parallel concatenated
ensemble of this section. It is interesting to note that for a fixed
local code, the serial construction turns out to be more restrictive 
that parallel.
In particular, \cite{bar01e} shows that serially concatenated codes
with a fixed inner code and random outer code approach the GV bound
only for rate $R\to 0$. They are also asymptotically good, although
below the GV bound, for a certain range of code rates depending
on the code $A$.

The results about the weight spectrum of bipartite-graph codes can
also be used to estimate the ensemble average error exponent of 
codes under maximum likelihood decoding.
This is a relatively standard calculation that can be performed
in several ways; we shall not dwell on the details here. Of course,
for code rates $R\le 0.202$ when the codes meet the GV bound and
their weight spectrum is binomial, the error exponent of their
maximum likelihood decoding will meet Gallager's bound $E_0(R,p)$.
Similar results were earlier established for serial concatenations
\cite{blo82,tho87}.

\section{Improved estimate of the distance}\label{sect:improved}
In this section we present a constructive family of one-level, parallel
concatenated codes that surpass the product bound on the distance
for all code rates $R\in (0,1).$

The intuition behind the analysis below is as follows.
The distance of two-level code constructions such as Forney's concatenated
codes and similar ones is often estimated by the product of the
distances of component codes. In (\ref{eq:product})
this result is established for expander codes (note that its proof,
different from the corresponding proofs for serial concatenations, is based
on the expanding properties of the graph $G_1$).

It has long been recognized that apart from some special cases (such as
product codes and the like) the actual relative minimum distance of two-level
codes often exceeds the relative ``designed distance'' which in this case is
the product $\delta_0\delta_1.$ To see why this is the case
let us recall the serial concatenated construction which is 
obtained from an $[n_1,k_1,d_1]$ $q$-ary Reed-Solomon
code $B$, $q=2^{k_0}$ 
and an $[n_0,k_0,d_0]$ binary code $A$.
A typical codeword of the concatenated code $C$ can be thought of as
a binary $n_0\times n_1$-matrix in which the $i$th column, $1\le i\le n_1,$
represents an encoding with the code $A$ of the binary representation
of the $i$th symbol of the codeword in $B$.

A codeword of weight $d_0d_1$ in the code $C$ can be obtained only if
there exists a codeword of weight $d_1$ in the code $B$ in which every
symbol is mapped on a codeword of weight $d_0$ in the code $A$.
By experience, the true distance of the code $C$ exceeds the product bound
substantially (for instance,
on the average concatenated codes approach the GV distance;
see the end of Sect.~\ref{sect:random} ), although quantifying this
phenomenon for constructive code families is a difficult problem.

The situation is different for expander codes (we will analyze the
modified construction of the previous section) because the component
codes $A$ and $B$ are of constant length, so we can have more control of
both the binary and the $q$-ary weight of the symbols in the codeword
and still obtain a constructive code family. The analysis below is
based on the following intuition: the codes $A$ and $B$'s roles are
not symmetric. If the product bound $\delta_0\delta_1$ were to be achieved
by some codeword of $C$,
then the subcodewords corresponding to vertices of $V_1$ would have
a relatively low $q$-ary weight (equal to $\delta_1$) but a relatively
high binary weight, concentrated into few $q$-ary symbols. On the
other hand the subcodewords corresponding to vertices of $V_0$ would
spread out their binary weight among all their $q$-ary symbols each of
which would have a relatively low binary weight. The edges of the
bipartite graph correspond to symbols of the two codes, making these
conditions incompatible.

We now elaborate on this idea, beginning with the code construction
of Sect. \ref{sect:mult}. The analysis in this case is simple
and paves way for a more complicated calculation for the modified
bipartite-graph codes and an improved distance bound.

\subsection{Basic construction} \label{sec:4.1}
Let us estimate the minimum binary weight of a codeword in the code
$C(G;A,B)$ where $G$ is a graph with a small second eigenvalue
$\lambda$. Recall that $E(v)$ denotes the
set of edges incident on a vertex $v$.

Let us introduce some notation.
Let $\bfx\in C$ be a codeword. For a given vertex $v$,
the subvector $\bfx_v \in \{0,1\}^{\Delta t}$ can 
be partitioned into $\Delta$ consecutive segments of $t$ bits,
We write $\bfx=(\bfx_1,\dots, \bfx_\Delta),$ where
   $$
     \bfx_i=(x_{t(i-1)+j},1\le j\le t), \quad 1\le i\le \Delta,
   $$
 each segment corresponding to its own edge $e\in E(v)$.
The Hamming weight $\wt(\bfx_i)$ will be also called the binary weight 
of the edge $e$, denoted $w_b(e)$. The corresponding relative weight
of the edge is denoted by $\omega_b(e)=\wt_b(e)/t.$
We call an edge {\em nonzero} relative to the codeword $\bfx$ if 
$\wt_b(e)\ne0.$ The number of nonzero edges of $\bfx$ is called
the $q$-ary weight of $\bfx.$

For a subset of vertices
$S\subset V_i, i=1,2$ let $E(S)=\cup_{v\in S} E(v).$ 
For two subsets $S\in V_0, T\in V_1$
denote by $G_{S\cup T}$ the subgraph of $G$ induced by $S$ and $T$
and let $(S,T)$ be the set of its edges. In particular, if $T$ is just
one vertex, we denote by $(S,v)$ the set of edges that connect $v$ and
$S$. Let $\deg_S(v)=|(S,v)|.$

Consider a codeword $\bfx$ of the code $C.$ Let $S\subset V_0$
be the smallest subset of left vertices that contains all the
nonzero coordinates of $\bfx,$ and let $T\subset V_1$ be the same
for right vertices. Formally, $\supp(\bfx)\subset(S,T),$
and both $S$ and $T$ are minimum subsets by inclusion that satisfy
this property. Note that all edges in $G\backslash G_{S\cup T}$
correspond to zero symbols of $\bfx$ (but there may be additional zero
symbols).

Let $\gamma=\gamma(\bfx)$ be the average, over all edges 
$e$ that join a vertex of $S$ to a vertex of $T$
of the relative binary weight of $e$:
 \begin{equation}
     \gamma=\frac 1{|(S,T)|} \sum_{e\in (S, T)} \omega_b(e).
           \label{eq:gamma}
   \end{equation}
Let $v$ be some vertex, either of $S$ or of $T$. Let us define two local 
parameters $\beta_v$, $\gamma_v$. 
These parameters are relative to the codeword $\bfx$. 
\begin{itemize}
\item
The quantity $\beta_v$ is defined as the average, over all
{\em non-zero} edges $e$ incident to $v$, of the relative (to $t$) 
 binary weight $\omega_b(e)$:
\item The quantity $\gamma_v$ is defined as the average, over all 
edges $e$, zero or not,
  \begin{itemize}
  \item that join $v$ to a vertex of $T$ if $v\in S$,
  \item that join $v$ to a vertex of $S$ if $v\in T$,
  \end{itemize}
of the relative binary weight $\omega_b(e)$ of $e$. 
For instance, if $v\in T$, then
$$\gamma_v=\frac{1}{\deg_S(v)}\sum_{e\in (S,v
)}\omega_b(e).$$
Note that $\gamma_v\le \beta_v.$ 
\end{itemize}

We will use the big-O and little-o notation 
relative to
functions of the degree $\Delta.$ For instance $\bigo(1/\sqrt\Delta)$
denotes a quantity bounded above by $c/\Delta$, where $c$ does not
depend on $\Delta.$

Before we proceed we need to recall the following ``expander mixing'' lemma:

\begin{lemma}\label{lem:expander}
Let $G=(V_0\cup V_1, E)$ be a $\Delta$-regular
bipartite graph, $|V_0|=|V_1|=n$, with second eigenvalue $\lambda.$
Let $S\subset V_0,\; |S|=\sigma n.$ 
Let $\alpha>\lambda/2\sigma\Delta.$
Let $U\subset V_1$ be 
defined by $U=\{v\in V_1: \deg_S(v)\ge (1+\alpha)\sigma\Delta\},$ then
  $$
    |U|\le \frac{\lambda}{2\sigma\Delta\alpha-\lambda}|S|.
  $$
\end{lemma}
Below we assume that $G$ is a Ramanujan graph implying that $\lambda\le
2\sqrt{\Delta-1}.$ Recall from Lemma \ref{lemma:expanding} that since
$\delta_0$ and $\delta_1$ are fixed, the value $\sigma$ is
lowerbounded by a quantity independent of $\Delta$ and can be thought
of as a constant in the following analysis.

Using this in the above lemma, we obtain
  \begin{equation}\label{eq:sqrt}
   |U|\le \frac{\lambda\sigma n}{2\sigma\Delta\alpha-\lambda}
    \le \frac{\sigma n}{\sigma\alpha\sqrt{\Delta}-1}=\frac{cn}{\sqrt{\Delta}},
  \end{equation}
where $c=c(\sigma,\alpha)=1/(\alpha-(1/\sigma\sqrt{\Delta})).$
We will choose $\alpha$ to be a quantity that, when $\Delta$
grows, tends to zero and is such that $\alpha\sqrt{\Delta}$ tends to
$\infty$: what (\ref{eq:sqrt}) shows us is that $|U|/n$ is a vanishing
quantity when $\Delta$ grows, which we will write as 
$|U|/n = o_\Delta(1)$.
Similarly, applying Lemma \ref{lem:expander} in the same way
to the set $\bar S=V_0\backslash S,$
we obtain the following corollary.
\begin{corollary}\label{cor} 
Let $\alpha$ be such that 
$\alpha =o_\Delta(1)$ and $1/\alpha\sqrt{\Delta}=o_\Delta(1)$.
Let 
$$R_\alpha=\{v\in V_1:(1-\alpha)\sigma \Delta\leq \deg_S(v)\le
(1+\alpha)\sigma \Delta\}.$$ Then $1-|R_\alpha|/n = o_\Delta(1)$.
\end{corollary}

What the expander lemma essentially says is that for any set $S$ of vertices
of $V_0$, almost every vertex of $V_1$ will have a proportion of its edges
incident to $S$ that almost equals $\sigma =|S|/n$. 
Going back to the sets $S$ and $T$ associated to the codeword $\bfx$,
the consequence of this
is that $\gamma$ is essentially obtained by simple averaging of
the $\gamma_v$'s: more precisely,

\begin{lemma}\label{lem:gamma} 
  $$\gamma = \frac{1}{|S|}\sum_{v\in S}\gamma_v +o_\Delta(1)
           = \frac{1}{|T|}\sum_{v\in T}\gamma_v +o_\Delta(1).$$
\end{lemma}
\begin{proof}
For instance, let us prove the second equality. 
Let $|S|=\sigma n, |T|=\tau n.$ 

First write 
  \begin{align}
  |(S,T)| & = \sum_{v\in T}\deg_S(v) = 
              \sum_{v\in T\cap R_\alpha}\deg_S(v) + 
              \sum_{T\setminus R_\alpha}\deg_S(v)\nonumber\\
          & = \sum_{v\in T\cap R_\alpha}\Delta(\sigma +o_\Delta(1)) 
              + n\Delta o_\Delta(1)\nonumber
  \end{align}
by Corollary \ref{cor}. We obtain, again by Corollary \ref{cor},
\begin{equation}
  \label{eq:(S,T)}
  \frac{|(S,T)|}{n\Delta} = \sigma\tau + o_\Delta(1).
\end{equation}
Next, by definition of $\gamma$ and $\gamma_v$,
  $$|(S,T)|\gamma = \sum_{v\in T}\deg_S(v)\gamma_v.$$
As above, partition $T$ into $T\cap R_\alpha$ and $T\setminus R_\alpha$
and apply Corollary \ref{cor} to obtain
  \begin{align}
 |(S,T)|\gamma & = \sum_{v\in T\cap R_\alpha}\sigma\Delta\gamma_v 
                   + n\Delta o_\Delta(1)\nonumber\\
               & = \sum_{v\in T}\sigma\Delta\gamma_v 
                   + n\Delta o_\Delta(1).\label{eq:(S,T)gamma}
  \end{align}
Now rewriting (\ref{eq:(S,T)}) as $|(S,T)| =|T|\sigma\Delta(1+o_\Delta(1))$
and dividing it out of (\ref{eq:(S,T)gamma}) gives the result.
\end{proof}

Our strategy will be to consider $\gamma$ as a parameter liable to 
vary between $0$ and $1$. For every possible $\gamma$ we shall find
a lower bound for the total weight $\delta(\gamma)$
of $\bfx$ and then minimize over $\gamma$.
We have introduced the two local parameters 
$\beta_v$ and $\gamma_v$ for a technical reason: the quantity
$\beta_v$ is the natural one to consider when estimating the weight
of the local code at vertex $v$. However averaging the $\beta_v$'s when
$v$ ranges over $S$ or $T$ is tricky while Lemma \ref{lem:gamma}
enables us to manage the averaging of the $\gamma_v$ conveniently.

Now we introduce the
{\em constrained distance} of $A$: 
it is defined to be any function $\delta_0(\beta)$ of $\beta\in(0,1)$ that 
\begin{itemize}
  \item is $\cup$-{\em convex}, continuous for 
     $\beta$ bounded away from the ends of the interval,
          and is non-decreasing in $\beta,$
  \item is a lower bound on the minimum relative binary weight of a 
   codeword of $A$ under the restriction that the average binary weight 
   of its nonzero edges is equal to $\beta t$. 
 \end{itemize}
The next lemma should explain the purpose of this definition.
\begin{lemma}\label{lem:fixed}
Let $\bfx$ be some codeword of $C$ and let $S, |S|=\sigma n$ and $\gamma$ be
the quantities defined above. 
The binary weight $w_b(\bfx)=\omega(\bfx)N$ satisfies
   $$ \omega(\bfx) \geq \sigma\delta_0(\gamma) +o(1).$$
\end{lemma}

\noindent
{\em Proof :}
  We clearly have 
  $$\omega(\bfx) \geq \frac{\sigma}{|S|}\sum_{v\in S}\delta_0(\beta_v).$$
Now notice that by their definition $\beta_v\geq\gamma_v$ so that
$\delta_0(\beta_v)\geq\delta_0(\gamma_v)$ since 
$\delta_0(\cdot)$ is non-decreasing.
Furthermore, by convexity and uniform continuity
of $\delta_0$ and by Lemma \ref{lem:gamma},
\begin{equation*}
  \frac{1}{|S|}\sum_{v\in S}\delta_0(\gamma_v)\geq \delta_0(\gamma) +o(1).
  \end{equation*}
\vskip-5mm\hfill{\rule{1mm}{2mm}}

\bigskip
Next we bound $\sigma$ from below as a function of $\gamma$. We do this
in two steps. The first step is to evaluate a {\em constrained distance} 
$\delta_1(\beta)$ for $B$ defined as the minimum relative $q$-ary weight
of any nonzero codeword of  
$B$ such that the average binary weight of its nonzero symbols (edges) 
is equal to~$\beta t$.

The following lemma is an existence result obtained by the 
random choice method, 
but since the code $B$ is of fixed size, it can be chosen through 
exhaustive search without compromising constructibility.

\begin{lemma}\label{lem:B}
  For any $\varepsilon >0$,  and $t$ and $\Delta$ large enough,
  there exist codes $B$ of rate $R_1$ such that for any $0<\beta <1$,
   the minimum relative $\beta$-constrained $q$-ary weight $\delta_1(\beta)$ 
  of $B$ satisfies
  $$\delta_1(\beta) \geq \frac{1-R_1}{h(\beta)} -\varepsilon.$$
\end{lemma}
\begin{proof}
We use random choice analysis: 
let us count the number $N_w$ of vectors
  $\bfz\in\{0,1\}^{t\Delta}$ of $q$-ary weight $w=\omega \Delta$ such that the
average binary weight of its non-zero $q$-ary symbols is $\beta t$. 
Let $w_i,i=1,\dots, w$ be the weights of these non-zero $t$-tuples.
We have:
  $$
    N_w\le \binom \Delta w \sum_{\sum w_i= 
             w\beta t}\prod_{i=1}^w\binom t{w_i}.
  $$
By convexity of entropy, for sufficiently large $\Delta $ and $t$,
the largest term on right-hand side is when
all the $w_i$ are equal. Then
  $$
    N_w\lesssim \binom \Delta w \binom t {\beta t}^{w}\lesssim
     2^{\omega\Delta t h(\beta)}
  $$
when $t$ is large enough. 
Hence, for a randomly chosen code of rate $R_1$,
 the number $A_{\omega,\beta}$
of $\beta$-constrained codewords of relative weight
$\omega$ has an expected value
  $$
 \bar A_{\omega,\beta}\lesssim 2^{\Delta t(R_1-1+\omega h(\beta))}.
  $$
As long as $\omega$ is chosen so that the above exponent is less than zero,
there exists a code whose $\beta$-constrained minimum distance is at least
$\omega$: furthermore, since the number of
possible values of $(\omega ,\beta)$ (for which $w$ and $w\beta t$ are
integers) is not more than polynomial in $t\Delta$, we obtain the
existence of codes that satisfy our claim for all values of $\beta$.
\end{proof}

{\em Comments\/}: 
For $\beta <\delta_{GV}(R_1)$ we obtain values of $\delta_1(\beta)$ that
are greater than $1$. This simply means that no $\beta$-constrained
codewords exist.

It follows from \cite{tho83} that the same bound on the
$\beta$-constrained distance can be obtained for Reed-Solomon codes
over $GF(2^t)$ whose symbols are mapped to binary $t$-vectors by
random linear transformations. Thus, it is possible to prove the
results of this section restricting oneself to Reed-Solomon $q$-ary
codes $B$. 

\medskip

From now on we assume that $B$ is chosen in the way guaranteed by 
Lemma~\ref{lem:B}.
We can now prove:
 
\begin{lemma}\label{lem:fixed2}
Let $\epsilon>0$. For a codeword $\bfx\in C,$ let $S$ and $\gamma$
be defined as in {\rm(\ref{eq:gamma})} and let $\sigma=|S|/n$.
There exist $\Delta$ and $t$ such that 
  $$\sigma \geq \frac{1-R_1}{\oh(\gamma)} -\varepsilon,$$
where $\oh(\beta)$ is defined as $\oh(\beta)=h(\beta)$ for
$0\leq\beta\leq 1/2$
and $\oh(\beta)=1$ for $1/2\leq \beta \leq 1$.
\end{lemma}

\begin{proof}
 By Lemma \ref{lem:gamma} we have $\sum_{v\in T}\gamma_v \le \gamma |T|-
\epsilon$.
 Therefore  there must be a possibly small
 but non-negligible subset of right vertices $T_1\subset T$ (namely at least
 $|T_1|\ge \varepsilon |T|$) for which $\gamma_v\leq \gamma +\varepsilon$. 
By Corollary~\ref{cor} the subset of vertices 
$T_2\subset T_1$ that
do not satisfy $(1-\alpha)\sigma\Delta \le \deg_S(v)\le (1+\alpha)\sigma
\Delta$ is of size $|T_2|=no_{\Delta}(1)$ for $\alpha$ arbitrarily
small. 
Consider a vertex $v\in T_1\backslash T_2$ 
and let $\omega_v$ be its relative $q$-ary weight. 
Let $\alpha'$ be the proportion of nonzero edges among the edges from
$v$ into $S$. Since $\deg_S(v)$ can be taken to be arbitrarily close
to $\sigma\Delta$ we write, dropping vanishing terms, 
$\alpha' = \omega_v/\sigma$.
By their definitions we have $\beta_v=\gamma_v/\alpha'$. 
By Lemma \ref{lem:B} we have
 $\omega_v \geq (1-R_1)/h(\beta_v)$ and therefore
  $$\sigma \geq \frac{1-R_1}{\alpha' h(\gamma_v/\alpha')}.$$
 But, noticing that the function $h$ is $\cap$-convex, we have
$h(x)\geq \alpha' h(x/\alpha')$
 for any $x$ and any $\alpha' \leq 1$, so that 
$\sigma\geq(1-R_1)/h(\gamma_v)$. Finally, we have $h(x)\leq \oh(x)$
for every $x$ and since $\gamma_v\leq \gamma$ (omitting $\varepsilon$
terms) and $\oh$ is
non-decreasing we have $h(\gamma_v)\leq \oh(\gamma)$ which proves the
result.
\end{proof}

Let us now estimate $\delta_0(\beta).$ 

\begin{lemma}\label{lemma:db0}  Let $\epsilon >0, 0<\beta<1$,
$\lambda(\beta)=\beta/h(\beta).$ For
sufficiently large $\Delta$ and $t$ there exists a code $A$
for which a suitable function $\delta_0(\beta)$ is given by
\begin{equation}\label{eq:d0}
    \delta_0(\beta)=(1-R_0)g(\beta),
  \end{equation}
where
\begin{itemize}
  \item $g(\beta)=\delta_{GV}(R_0)/(1-R_0)$ if $\beta\leq \delta_{GV}(R_0)$,
  \item $g(\beta)=\lambda(\beta)$ if $\delta_{GV}(R_0)\leq \beta$ and
        $R_0\le 0.284,$
  \item If $\delta_{GV}(R_0)\leq \beta$ and $0.284\le R_0\le 1,$
 \end{itemize}     
       \begin{align}
         g(\beta)&=
       \frac{a\beta+b} {\beta_1-\dgv(R_0)}\quad\dgv(R_0)\le\beta\le\beta_1
                  \label{eq:d01}\\[0mm]
         g(\beta)&=\lambda(\beta) \quad\beta_1\le\beta\le 1,
               \label{eq:d02}
      \end{align}
where $\beta_1$ is the largest root of
   $$
      h(\beta)\Big(\beta-h(\beta)\frac{\dgv(R_0)}{1-R_0}\Big)
                    =-(\beta-\dgv(R_0))\log(1-\beta),
    $$
  $$
     a={\lambda(\beta_1)-\lambda(\dgv(R_0))},
 \quad b={\lambda(\dgv(R_0))\beta_1-\lambda(\beta_1)\dgv(R_0)}
  $$        
\end{lemma}  
\begin{proof} 
We again apply random choice: more precisely, let $A$ be chosen to
have rate $R_0$ and satisfy Lemma~\ref{lem:B}. 

Lemma \ref{lem:B} applied to the code $A$ tells us that if $\beta t$
is  the average binary weight of the nonzero $q$-ary
symbols of some codeword, then this codeword must have $q$-ary weight
at least $\Delta(1-R_0)/h(\beta)$: for $\beta < \delta_{GV}(R_0)$ this
quantity is larger that $\Delta$, meaning that such a codeword doesn't
exist and we may choose any value we like for $\delta_0(\beta)$.
For $\beta \geq \delta_{GV}(R_0)$, 
since the total binary weight of the codeword equals $\beta
t$ times its $q$-ary weight we obtain that this codeword has total 
binary weight at least $t\Delta(1-R_0)\lambda(\beta)$.

Now the function $\lambda(\beta)=\beta/h(\beta)$ 
is convex for $0.197\le \beta<1.$
Thus if $R_0\le 1-h(0.197),$ we can define
$\delta_0(\beta)=\delta_{GV}(R_0)$
for $0\leq\beta\leq\delta_{GV}(R_0)$ and
$\delta_0(\beta)=\lambda(\beta)$ for $\delta_{GV}(R_0)\leq \beta\leq 1$.
For greater values of the rate $R_0$ we must replace the non-convex part of the
curve $\lambda(\beta)$ with some convex function such as a tangent
to this curve. This results in elementary but cumbersome calculations
which lead to the claim of the lemma.
\end{proof}

\begin{figure}[tH]
\epsfysize=7cm
\setlength{\unitlength}{1cm}
\begin{center}
\begin{picture}(5,8)
\put(-1.5,1){\epsffile{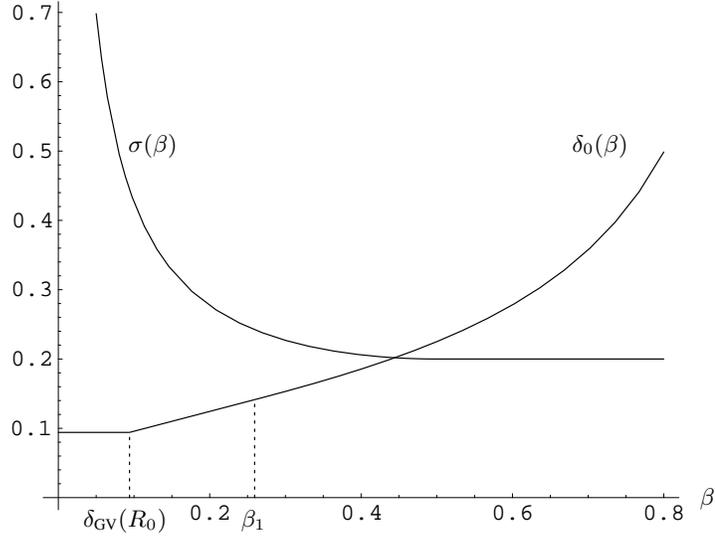}}
\put(-0.5,1){\mbox{\footnotesize$\dgv(R_0)$}}
\put(1.6,1){\mbox{\footnotesize$\beta_1$}}
\put(7.7,1.3){\mbox{\footnotesize$\beta$}}
\put(0.1,6){\mbox{\footnotesize$\sigma(\beta)$}}
\put(6,6){\mbox{\footnotesize$\delta_0(\beta)$}}
\end{picture}
\vskip-1cm
\caption{Estimates of the constrained distance of the 
code $A$ and of $\sigma$.}
\label{fig:d0d1}
\end{center}
\end{figure}

The behavior of the functions $\delta_0(\beta)$ and $\sigma=\sigma(\beta)$ is
sketched in Fig.~\ref{fig:d0d1}. 

\medskip

Now together, Lemmas \ref{lem:fixed}, \ref{lem:fixed2} and  
\ref{lemma:db0}
give us the following lower bound on the relative distance $\delta$
of the code $C(G;A,B)$: 
   $$
    \delta \ge 
        \min_{0\le\beta\le 1} (1-R_0)(1-R_1)\frac{g(\beta)}
                {\oh(\beta)}.
   $$
Since $g(\beta)$ is non-decreasing and $\oh(\beta)$ is constant
for $\beta \geq 1/2$, the minimum is clearly achieved
for $\beta \leq 1/2$: similarly, $\oh(\beta)$ is non-decreasing and
$g(\beta)$ is constant for $\beta\leq \dgv(R_0)$ so that the minimum
must be achieved for $\beta \geq \dgv(R_0)$.
We can therefore limit $\beta$ to the interval
$(\dgv(R_0),1/2)$ and replace $\oh(\beta)$ by $h(\beta)$. Optimizing
on $R_0$ to get the best possible $\delta$ for a given code rate $R$, we get:
   $$
    \delta \ge \max_{\begin{substack}{R_0,R_1\\ R_0-R\le 1-R_1}
          \end{substack}}
        \min_{\dgv(R_0)\le\beta\le 1/2} (1-R_0)(1-R_1)\frac{g(\beta)}
                {h(\beta)}.
   $$
The full optimization is possible only
numerically, but we can make one simplification which entails
only small changes in the value of $\delta(R)$. Namely, let us 
optimize on the rates of component codes $R_0,R_1$ ignoring the dependence
of $g(\beta)$ on $R_0$. Let us choose $R_0$ to satisfy $1-R_0=\frac12(1-R), 
$ then $1-R_1\ge \frac12(1-R),$ and we obtain the bound given in the
following theorem.
\begin{theorem} There exists an easily constructible family of
binary linear codes $C(G;A,B)$ of length $N=n\Delta,n\to\infty$ and rate $R$
whose relative distance satisfies
  \begin{equation}\label{eq:bb} 
   \delta(R)\ge \frac14 (1-R)^2\min_{\dgv(\frac{1+R}2)<\beta<1/2} 
      \frac {g(\beta)}
       {h(\beta)}-\epsilon \qquad(\epsilon>0),
   \end{equation}
where $g(\beta)$ is defined in (\ref{eq:d01})-(\ref{eq:d02}).
\end{theorem}
We remark that the bound (\ref{eq:bb}) is very close to (but below) the 
curve $0.0949(1-R)^2$ which can be used for rough comparison with
other bounds in this context.

Note that the overall complexity of finding the codes $A,B$ does not 
grow with $n$, so the complexity of constructing the code $C$ is 
proportional to the complexity of constructing the graph $G$.

\subsection{Modified construction}\label{sec:4.2}
Let us consider a modified expander code $C(G;A,B,A_{\aux})$ 
of Section \ref{sect:mod} which should be consulted for notation.
We wish to repeat the argument of the previous section. 

The quantities $\gamma$ and $\beta$ are defined as before, but on the
subgraph $G_1={V_0\cup V_1,E_1}.$ 
For $\beta\in (0,1)$ we again 
define $\delta_1(\beta)$ as the minimum relative $q$-ary weight
of any nonzero codeword of  
$B$ such that the average binary weight of its nonzero symbols (edges) 
is at most $\beta t$. Lemmas \ref{lem:fixed}, \ref{lem:B},
\ref{lem:fixed2} hold unchanged.
The definition of $\delta_0(\beta)$ is somewhat more complicated however,
because the weight of the check edges of the code $A$ is now unconstrained.
Let us fix an information set $I$ of $R_0\Delta$ 
$q$-ary symbols (for definiteness, suppose that they occupy the first 
coordinates of the codeword). Let $\delta_0(\beta)$ be 
\begin{itemize}
  \item a $\cup$-{\em convex} continuous function of $\beta$
  \item a lower bound the minimum relative binary weight of a 
   codeword of $A$ under the restriction that the average binary weight 
   of its nonzero edges {\bf in \em I}\/ is at least $\beta t$. 
 \end{itemize}

The following lemma gives an estimate for $\delta_0(\beta)$ that will
replace Lemma \ref{lemma:db0}.

\begin{lemma}\label{lemma:di}
For any $\epsilon >0$, 
  for $\Delta_1$ and $t$ large enough,
  for any $R_0$, there exist codes of rate $R_0$ such that, for every 
  $\beta$, $\delta_0(\beta)+\epsilon$ is greater than any convex function
  that does not exceed $\omega(\beta)$, where 
  $\omega(\beta)$ is the root of the equation in $\omega$
  $$
   1-R_0 = \max_{R_0\omega_1+(1-R_0)\omega_2=\omega}
     R_0\omega_1\frac{h(\beta)}{\beta} + (1-R_0)h(\omega_2),$$
  where $\omega$ and $\omega_1$ are constrained by
   $h^{-1}(1-R_0)\leq \omega\leq \beta$
  and $\omega_1\leq\beta$.
\end{lemma}
\begin{proof} Let $A$ be a linear code of rate $R_0$ and relative
distance $\dgv(R_0).$ Let $\bfx\in A$ be a codeword of weight
$w=\omega t\Delta$ such that its average binary weight of nonzero symbols 
in $I$ is $\beta t$. Denote by $\omega_1$ the proportion of nonzero
bits among the symbols (edges) of $I$ and let $\omega_2$ be the same
for $J=\{1,\dots,n\}\backslash I.$ 
Let us estimate the total number $M_w$ of such vectors $\bfx.$
As in previous proofs, we take the assumption that each nonzero symbol in $I$
if of weight exactly $\beta t$, justified by the fact that the entropy
function is $\cap$-convex. Hence the number of nonzero symbols in $I$
is 
   $$
   \frac{R_0\omega_1\Delta t}{\beta t}=\frac{R_0\Delta\omega_1}{\beta}.
   $$
With regard to the symbols in $J$ there
are no restrictions apart from their total weight. Hence
  $$
     M_w\approx \max_{R_0\omega_1+(1-R_0)\omega_2=\omega}
            \binom{(1-R_0)t\Delta}{\omega_2(1-R_0)t\Delta}
            (2^{th(\beta)})^{R_0\Delta\omega_1/\beta}.
  $$
Then, with respect to the ensemble of random linear binary
$[\Delta t,R_0\Delta t]$ codes, the probability that 
$\delta_0(\beta)\le \omega(\beta)$ is 
  $$
   \Pr[\wt(x)=\omega n]\lesssim 2^{-\Delta t(1-R_0)}M_w.
  $$
For any $\omega>\omega(\beta)$
this probability is less than 1, so there exist codes that satisfy
the claim of the lemma.
\end{proof}

\remove{The remaining part of the derivation deals with various technical issues
such as optimization on $\omega_1,\omega_2$ in Lemma \ref{lemma:di}
and dealing with 
(non)-convexity issues of the curve obtained, and optimizing
on $\beta$ and the choice of $R_0,R_1$ in Theorem }

\subsubsection*{Optimization on $\omega_1,\omega_2$ in Lemma \ref{lemma:di}.}
We need to maximize the function 
   $$
    F=R_0\omega_1\frac{h(\beta)}{\beta} + (1-R_0)h(\omega_2)
   $$
on $\omega_1,\omega_2$ under the condition 
$R_0\omega_1+(1-R_0)\omega_2=\omega.$
The maximum is attained for 
   $$
     \omega_1=\frac{1}{R_0}(\omega-(1-R_0)a(\beta)),\quad
   \omega_2=a(\beta),
   $$
where $a(\beta)=(2^{h(\beta)/\beta}+1)^{-1}.$
Substituting these values into the expression for $F$ and equating
the result to $1-R_0,$ we find the value of $\omega:$
  $$
    \omega=\omega^\ast(\beta):=
        (1-R_0)\Big[a(\beta)+\frac\beta{h(\beta)}(1-h(a(\beta)))\Big]
  $$
Next recall that $\omega,\omega_1$ are constrained as follows:
   $  \dgv(R_0)\le \omega\le \beta, \quad\beta\ge \omega_1.
   $
As it turns out, the unconstrained maximum computed above
contradicts these inequalities for values of $\beta$ close to $\dgv(R_0),$
namely, the value of $\omega$ falls below $\dgv(R_0).$
Therefore, define $\beta_1$ to be the (only) root of the equation in $\beta$
  $$
    \dgv(R_0)=\omega^\ast(\beta).
  $$
For $\beta\in [\dgv(R_0),\beta_1]$ let us take $\omega_1=\beta,
\omega_2=a(\beta).$ We then
use the condition $F=1-R_0$ to compute 
  $$
   \omega=\omega^{\ast\ast}(\beta):=
        R_0\beta+(1-R_0)h^{-1}\Big(1-\frac{R_0}{1-R_0}h(\beta)\Big).
  $$
Concluding, the value of $\omega(\beta)$ in Lemma \ref{lemma:di}
is given by
 \begin{equation*}
   \omega(\beta)=\begin{cases} \omega^{\ast\ast}(\beta) & 
               \dgv(R_0)\le \beta\le \beta_1\\
    \omega^\ast(\beta) &\beta_1\le \beta\le 1/2.
    \end{cases}
 \end{equation*}

By definition, the relative distance $\delta_0(\beta)$ is bounded
below by any convex function that does not exceed $\omega(\beta).$
The function $\omega(\beta)$ consists of two pieces, of which
$\omega^{\ast\ast}$ is a convex function but $\omega^\ast$ is not.
We then repeat the same argument as was given after Lemma \ref{lemma:db0},
replacing $\omega^\ast(\beta)$ with a tangent to $\omega^{\ast\ast}(\beta)$
drawn from the point $(1/2,\omega^\ast(1/2)).$ This finally gives the
sought bound on the function $\delta_0(\beta).$ We wish to spare the
reader the details.

\begin{figure}
\centerline{\psfig{figure=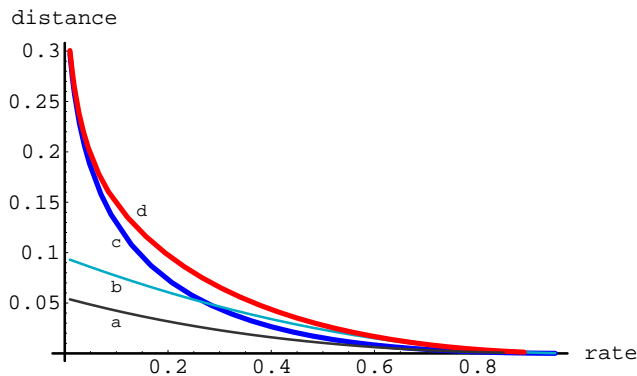,height=5cm}}
\caption{Parameters of expander codes: (a) Bound (\ref{eq:mult}),
(b) Bound (\ref{eq:bb}), 
(c) product (Zyablov) bound, (d) Bound (\ref{eq:ria}).}\label{fig:improved}
\end{figure}

\bigskip
The overall distance estimate follows from 
Lemmas \ref{lem:fixed}, \ref{lem:fixed2}, \ref{lem:B} and  
the expression for $\delta_0(\beta)$ found above. As before $\beta$
can be limited to the interval $(\dgv(R_0),1/2)$.
There is one essential difference compared
with the previous section: the rates $R_0,R_1$ of the component codes
are constrained by
(\ref{eq:Raux}) rather than (\ref{eq:rate}). Since $R_\aux$ is small,
essentially we have $R=R_0R_1.$ Thus we obtain the following result.
\begin{theorem}There exists an easily constructible family of
binary linear codes $C(G;A,B)$ of length $N=n\Delta,n\to\infty$
whose relative distance satisfies
  \begin{equation} \label{eq:ria}
    \delta(R)\ge \max_{R\le R_0\le 1}\;\min_{\dgv(R_0)<\beta<1/2} 
     \Big\{\delta_0(\beta,R_0)
        \frac{1-R/R_0}{h(\beta)}\Big\}-\epsilon.
  \end{equation}
\end{theorem}
Let us compare the distance estimates derived in this theorem and
in (\ref{eq:bb}) 
with the product
bound $\dgv(R_0)\dgv(R_1)$ on the code distance. Taking account of the
fact that the code $B$ is over the alphabet of size $q=2^t$ and
that for sufficiently large $t$, the bound $\dgv(R_1)$ comes arbitrarily
close to $1-R_1$, we obtain the (Zyablov) bound (\ref{eq:zyab})
   $$
\delta_{\text{Z}}(R)\ge \max_{R\le R_0\le 1}\dgv(R_0)(1-R/R_0).
   $$
In the table below we compute this bound and the new results
obtained in this paper. The bounds are also shown in Figure \ref{fig:improved}.

\bigskip\begin{tabular}{cccccccccc}
$R$ &0.1 &0.2 &0.3&0.4&0.5&0.6&0.7&0.8&0.9 \\
$\delta_{\text{Z}}(R)$ &0.129 &0.073 &0.044&0.026&0.015 &0.008&0.0040
&0.0015&0.00030\\ 
improved bound (\ref{eq:bb})&0.077 &0.061 &0.046 &0.034 &0.024 &0.015 &0.0084 &0.0037
 &0.00089\\
improved bound (\ref{eq:ria})&0.148 &0.095 &0.063 &0.041 &0.026
&0.015 &0.0078 &0.0031 &0.00073
\end{tabular}.

\bigskip
It is interesting that an improvement of the product bound is obtained
already with the basic construction of Sect.~\ref{sec:4.1}. Moreover,
for large rates this construction gives codes with a distance larger
than of the modified bipartite-graph construction of Sect.~\ref{sec:4.2}.
On the other hand, the modified construction asymptotically
improves the product bound for all values of the code rate 
other than $0$ and $1$. Both code families are polynomially constructible.

Concluding this section we remark that in principle,
the techniques presented here will generalize
to yield an improvement of the
bound \cite{bil04} for codes from hypergraphs.

\section{Families of asymptotically good
binary codes and their construction complexity}\label{sect:complexity}
Here we compare the parameters and construction complexity of other 
asymptotically good 
families of binary codes with the families of bipartite-graph codes 
presented in the previous section. 

First note that the complexity of specifying the bipartite-graph codes
is proportional to the complexity of describing the graph $G$ which
is $O(N\log N)$ if the graph is represented by a permutation of the
$N$ vertices in any one part. For most Ramanujan graphs, constructing
this permutation has complexity not more than $O(N\log N)$.
The complexity of constructing codes meeting the Zyablov 
bound (\ref{eq:zyab}) in the traditional way
is at most $O(N^2)$, by a combination of the
results in \cite{zya71,she93}.

Codes of distance greater than that
given by the Zyablov bound can be constructed
as multilevel concatenations \cite{blo82} or as concatenations
of good binary codes of relatively small length with algebraic
geometry codes from asymptotically maximal curves \cite{kat84}.
The parameters of multilevel concatenations of order $m$ are given by the
following Blokh-Zyablov bounds \cite{blo82}:
  $$
    R^{(m)}(\delta)=\max_{R_0\le 1-h(\delta)} \Big\{R_0-\frac{\delta R_0}m
          \sum_{i=1}^m\Big[\dgv\Big(R_0\frac{m-i+1}m\Big)^{-1}\Big]\Big\},
       \quad m=1,2,\dots .
  $$
(in this case it is more convenient 
to specify the code rate $R$ for a fixed value of the relative distance
$\delta$). Note that for $m=1$ we again obtain (\ref{eq:zyab}).
The value $R^{(m)}(\delta)$ 
increases monotonically with $m$ for any $0<\delta<1/2$
and thus these codes surpass the Zyablov bound for all $m\ge 2.$
Their construction complexity is $O(N^{\max(2,(m/R)-1)})$
which is higher than the complexity of constructing the bipartite-graph
codes, particularly for low code rates. The bound $R^{(m)}(\delta)$ is better than the value of the rate 
obtained for bipartite-graph codes beginning with $m=4$ or so.

The largest code rate of
multilevel concatenations is obtained by letting $m\to\infty$. The resulting
bound is given by
  $$
    R^{(\infty)}(\delta)=1-h(\delta)-\delta\int_0^{1-h(\delta)}\frac {dx}{\dgv(x)}
  $$
(this expression is called the {\em Blokh-Zyablov bound}).

Concatenations of algebraic-geometry codes with short binary codes
introduced in \cite{kat84}
improve the Blokh-Zyablov bound $R^{(\infty)}$ for all $0<\delta<1/2$
(but do not meet the GV bound on the ensemble-average distance of
concatenated codes). The code rate of these codes is the largest known
asymptotically (for a given $\delta$)
among families of binary codes with polynomial
construction complexity.
The construction complexity of this code family is $O(N^3\log^3 N)$
by a recent result of K.~Shum et al.~\cite{shu01}. 

Perhaps the most important is the fact that even though the two families
mentioned above have better parameters than  bipartite-graph codes,
their designed distance in both cases is estimated by the product bound.
On the contrary, the code families constructed in this paper 
provide an asymptotic improvement of the product bound on the distance.

\renewcommand\baselinestretch{0.9}
{\footnotesize
\providecommand{\bysame}{\leavevmode\hbox to3em{\hrulefill}\thinspace}

}
\end{document}